\pdfoutput=1

\documentclass[11pt]{article}

\usepackage[final]{acl}

\usepackage{times}
\usepackage{latexsym}
\usepackage{amsmath}
\usepackage{color}
\usepackage{CJKutf8}
\usepackage{makecell}
\usepackage{booktabs}
\usepackage{url}
\usepackage{multirow}
\usepackage{multicol}
\usepackage{makecell}
\usepackage[T1]{fontenc}

\usepackage[utf8]{inputenc}

\usepackage{microtype}

\usepackage{inconsolata}

\usepackage{graphicx}

\title{Real-time Ad retrieval via LLM-generative Commercial Intention \\
for Sponsored Search Advertising}

\author{
  \textbf{Tongtong Liu\textsuperscript{1}},
  \textbf{Zhaohui Wang\textsuperscript{1}},
  \textbf{Meiyue Qin\textsuperscript{1}},
  \textbf{Zenghui Lu\textsuperscript{1}},
  \\
  \textbf{Xudong Chen\textsuperscript{1}},
  \textbf{Yuekui Yang\textsuperscript{1}},
  \textbf{Peng Shu\textsuperscript{1}},
\\
\\
  \textsuperscript{1}Tencent Inc.
\\
  \small{
   \href{mailto:email@domain}{lttsmnliu@gmail.com}, \href{mailto:email@domain}{uniqueliu@tencent.com}
  }
}
\begin{document}
\maketitle
\begin{abstract}
The integration of Large Language Models (LLMs) with retrieval systems has shown promising potential in retrieving documents (docs) or advertisements (ads) for a given query. Existing LLM-based retrieval methods generate numeric or content-based DocIDs to retrieve docs/ads. However, the one-to-few mapping between numeric IDs and docs, along with the time-consuming content extraction, leads to semantic inefficiency and limits scalability in large-scale corpora. In this paper, we propose the \textbf{R}eal-time \textbf{A}d \textbf{RE}trieval (RARE) framework, which leverages LLM-generated text called Commercial Intentions (CIs) as an intermediate semantic representation to directly retrieve ads for queries in real-time. These CIs are generated by a customized LLM injected with commercial knowledge, enhancing its domain relevance. Each CI corresponds to multiple ads, yielding a lightweight and scalable set of CIs. RARE has been implemented in a real-world online system, handling daily search volumes in the hundreds of millions. The online implementation has yielded significant benefits: a 5.04\% increase in consumption, a 6.37\% rise in  Gross Merchandise Volume (GMV), a 1.28\% enhancement in click-through rate (CTR) and a 5.29\% increase in shallow conversions. Extensive offline experiments show RARE's superiority over ten competitive baselines in four major categories.

\end{abstract}

\section{Introduction}

An advertising system is a commercial application designed to generate revenue by presenting targeted ads to users, primarily consisting of two modules: ad retrieval and ranking. As a crucial component, ad retrieval swiftly filters relevant advertisements from vast libraries containing millions or even billions of candidates in response to user queries. Traditional ad retrieval models follow a two-stage process (\citealp{baidu:99}, \citealp{TF-IDF:12}, \citealp{DSSM:15}), first retrieving keywords from queries and then using those keywords to fetch ads. 
However, existing two-stage retrieval methods amplify the difference between user queries and manually chosen keywords, resulting in numerous missed retrieval issues. The query-ad single-stage approach (\citealp{gong2023full}, \citealp{DR:18}) addresses missed recall by directly retrieving ads but still struggles with understanding deeper commercial intentions due to limited reasoning capabilities and domain knowledge.

In recent years, LLMs~\cite{zhao2023survey} have garnered widespread attention and made remarkable achievements in the fields of search and recommendation ~\cite{pradeep2023does,tang2024listwise,shi2025know}. Most LLM-based retrieval methods~\cite{lin2025can} first create an index of docs by training the model to link docs with their identifiers (DocIDs). During retrieval, the model processes a query and generates the corresponding DocIDs~\cite{li2024matching}. For example, DSI (\citealp{Dsi:24}) employs numeric IDs to represent documents and establish connections between user queries and numeric IDs. LTRGR~\cite{learning:103} extracts document content, i.e., article title and body, to represent the document and implement the retrieval from a user query to a document.

Using heavy DocIDs~\cite{zeng2023scalable} presents several \textbf{drawbacks}. Firstly, the inference efficiency is low due to the one-to-few mapping between DocIDs and candidates~\cite{wang2024enhanced}, making it difficult to achieve real-time generation in large-scale scenarios. 
Secondly, representing docs/ads solely with heavy DocIDs fails to fully leverage the capabilities of LLMs in commercial intent mining and their advanced text generation abilities, thus hindering the effective exploration of the advertiser's intent.
Thirdly, it exhibits poor generalization. When new candidates emerge, it often requires retraining the model or updating the FM-index~\cite{fmindex} to accommodate their DocIDs, making it difficult to quickly update or remove candidates. Due to the requirement for real-time fetching of large sets of ads aligned with the user's commercial intent in ad retrieval, the existing semantically inefficient DocIDs are impractical and unsuitable for the task. Therefore, leveraging the powerful semantic capabilities of LLMs to design more effective semantic tokens for indexing, along with developing a more comprehensive end-to-end architecture, has become a crucial challenge.

To address this challenge, we developed a real-time LLM-generative ad retrieval framework named RARE. This framework utilizes LLM-generated commercial intentions (CIs) as an intermediate semantic representation to directly connect queries to ads, rather than relying on manually chosen keywords or heavy document identifiers. Specifically, RARE initially utilizes a knowledge-injected LLM (offline) to generate CIs for the ads in the corpus. It then selects a limited but comprehensive set of CIs and constructs a \textbf{dynamic index} that maps these CIs to their corresponding ads in a one-to-many relationship. Upon receiving a query, the RARE uses customized LLM (online) to generate CIs in real-time and retrieves the corresponding ads from the pre-built index.

A key \textbf{innovation} of RARE lies in utilizing CIs generated by a customized LLM to serve as intermediate semantic DocIDs for linking query and ads. Customized LLM is developed by knowledge injection and format fine-tuning of the base LLM. Knowledge injection involves incorporating domain-specific information to enhance expertise in the advertising domain. Format fine-tuning ensures that the LLM outputs only CIs and improves decoding efficiency.
CIs are defined as aggregations of keywords, generated by the customized LLM based on relevant materials of ads. Compared to existing carefully designed DocIDs, CIs fully leverages the text generation capabilities of LLMs. The one-to-many correspondence between CIs and ads makes the decoding process highly efficient. For new ads, RARE can generates CIs with the technique of constrained beam search, without the need to retrain the mode. Keyword bidding in the traditional query-keyword-ads paradigm introduces the possibility of index manipulation. In contrast to keywords, CIs are generated by LLMs equipped with world knowledge and commercial expertise, allowing for a better exploration of the commercial intent behind ads and queries.

The main contributions of our work are as follows:
(1) We propose a novel end-to-end generative retrieval framework named RARE to achieve real-time retrieval, which is the first known work on LLM-generative architecture that displays real-time retrieving on millions of databases. (2) We propose a method for knowledge injection and format fine-tuning to enable the base LLM to uncover the deep commercial intentions of advertisers and users, which are expressed as CIs. (3) We have deployed an online system based on LLMs for real-time inference and ad retrieval, which serves tens of millions of users in real-world scenarios everyday. (4) We conduct online A/B testing and offline experiments to verify the effectiveness of RARE.  A/B testing has yielded a 5.04\% increase in consumption, a 6.37\% increase in Gross Merchandise Value (GMV), a 1.28\% increase in Click-Through Rate (CTR), a 5.29\% increase in shallow conversions, and a remarkable 24.77\% increase in deep conversions. Simultaneously, in terms of offline evaluation metrics, RARE demonstrates superior performance in HR@500, MAP, and ACR metrics compared to 10 other competitive baselines.

\begin{figure*}[t]
  \centerline{\includegraphics[width=0.90\linewidth]{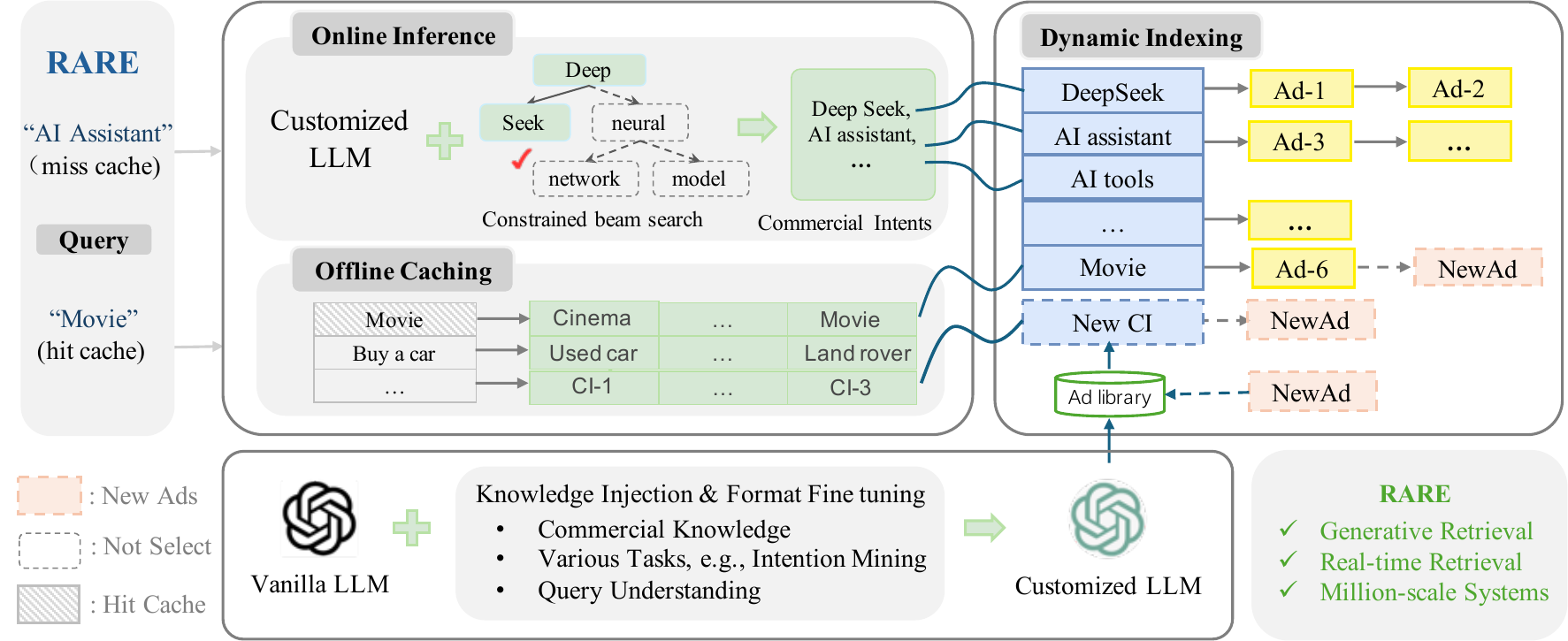}}
  \caption { The Real-time LLM-Generative Ad Retrieval framework (RARE) processes user queries by generating commercial intentions (CIs) through  LLM/caching, which are subsequently used to retrieve ads from the dynamic index. The customized LLM are created by injecting knowledge and learning rules based on vanilla LLM.
  \label{fig:LOGR}
}
\end{figure*}

\section{Related Works}

\paragraph{Ad Retrieval.} Traditional ad retrieval (\citealp{zhao2024survey}, \citealp{wang2024scaling}) typically follows a query-keyword-ad architecture, where queries retrieve keywords that are then used to pull ads. This approach includes both word-based and semantic-based methods. Word-based methods (\citealp{TF-IDF:12}, \citealp{bm25}) parse user queries to obtain keywords and use an inverted index to retrieve candidate ads. Semantic-based methods (\citealp{TF-IDF:12},\citealp{pretrained}) use a dual encoder to obtain embeddings for queries and keywords in a shared semantic space, enabling retrieval based on semantic similarities. These methods rely on manually chosen keywords resulting in numerous missed retrieval issues. In contrast, generation-based LLM retrieval methods (\citealp{sun2024learning}, \citealp{lin2024data}, \citealp{tang2024self}) uses DocIDs to represent ads, with the LLM directly generating the DocID corresponding to the candidate ads upon receiving a query.

\paragraph{Generative-based LLMs Retriever.} Generative based retrievers utilize the generative capabilities of LLMs to construct end-to-end retrieval models. Some approaches, such as DSI~\cite{Dsi:24}, NCI~\cite{nci:100}, Tiger~\cite{Tiger:23}, use document IDs as the generation target to implement query retrieval for docs/ads. These methods leverage LLMs to learn the correspondence between docs/ads and their IDs, directly generating the ID of the relevant docs/ads for query retrieval.  Other approaches, such as SEAL~\cite{autoregressive:101}and LTRGR~\cite{learning:103}, use document content as an intermediary to achieve document retrieval. They employ FM-Index to generate fragments that appear in the document, facilitating query-to-document retrieval. MINDER~\cite{multiview:102} employs pseudo-queries and document content for retrieval, but this significantly increases indexing volume, making it unsuitable for scenarios with large candidate sets.

\paragraph{Semantic DocIDs.}  LLM-generative retrieval typically employs DocIDs to perform query-to-document retrieval tasks. Existing DocIDs mainly include numeric IDs and document content. For instance, the numeric IDs in Tiger is represented as a tuple of discrete semantic tokens. In LTRGR, document content consists of predefined sequences that appear within the document. However, the semantic tokens used in these approaches are ID-like features, which suffer from low decoding efficiency since each DocID corresponds to few candidates. For new candidate docs or ads, it is necessary to retrain the model or rebuild the FM-Index to obtain their DocIDs, making it challenging to fast update or delete ads.

\section{Method}
\begin{figure}
  \includegraphics[width=0.9\columnwidth]{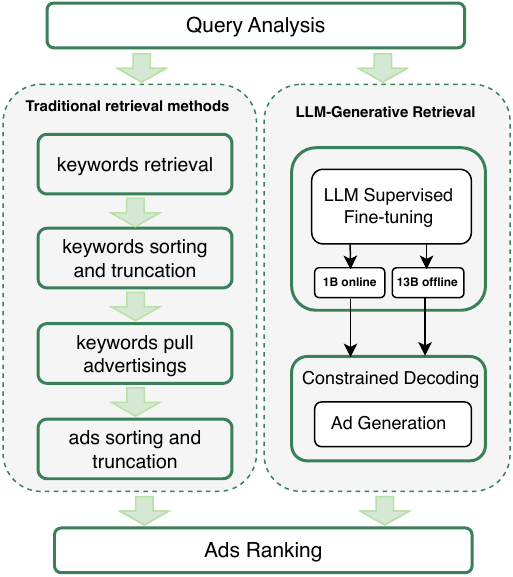}
  \caption{Comparison of RARE and Traditional Retrieval Methods. The Direct Generation of Candidate Ads from User Queries Shortens Link Structure.}
  \label{fig:end2end}
\end{figure}

In this paper, we introduce a novel end-to-end generative retrieval architecture designed for online retrieval, named \textbf{R}eal-time \textbf{A}d \textbf{re}trieval (RARE). RARE effectively shortens the link structure, which allows advertisements to overcome the limitations of keyword bidding and helps advertisers acquire more accurate traffic, as illustrated in Figure~\ref{fig:end2end}. 

\subsection{An End-to-end Generative Architecture}

Upon receiving a user query, RARE first analyzes it to generate corresponding Commercial Intents (CIs)\textemdash text with specific linguistic meaning\textemdash and then utilizes these CIs to retrieve the final ads. In the following, we detail the indexing of CIs to ads and explain the retrieval process.

\paragraph{Indexing.} RARE first generates CIs for the entire ad corpus and determines the commercial intention set, then the inverted index of CIs-Ads are built. For subsequent new ads, we perform constrained inference based on the current commercial intention set to ensure that each new candidate can be accurately updated in the index. Notably, CIs are texts with specific linguistic meanings generated by customized LLM to mine the commercial intention of ads. Further details on the implementation are discussed in Section \ref{sec:indexing}.

\paragraph{Retrieval.} The real-time generation of CIs for queries is based on a combination of offline caching strategies and online inference. The inferred CIs of high-frequency queries are stored in the cache. When a query arrives, RARE first checks whether the current query matches an entry in the cache. If a match is found, RARE directly retrieves the corresponding CIs to fetch ads. Otherwise, RARE uses the customized LLM with constrained beam search for real-time inference. The detailed implementations are introduced in \ref{sec:reasoning}.

The traditional query-keyword-ads architecture utilizes manually purchased keywords as search targets, subsequently retrieving ads based on a \textbf{fixed/predetermined index} linking keywords to ads. In contrast, our RARE framework uses advertisements themselves as retrieval targets and utilizes CIs as a \textbf{dynamic bridge} to index these ads, enhancing the system's flexibility and accuracy. CIs, generated by LLMs using comprehensive information from ads/queries, facilitate the generation of high-quality ad candidates and deeper user intent modeling.

\subsection{Customized LLM} 
To enhance the LLM's understanding of commercial and advertising knowledge and to generate more accurate CIs, we performed knowledge injection into the base LLM. To achieve real-time inference, where the model directly outputs CIs based on the query without intermediate reasoning process, we conducted format fine-tuning on the LLM. Details on the customization of LLM and the data organization are present in Appendix ~\ref{appendixA}.

\paragraph{Stage 1: Knowledge Injection.} 

This stage primarily involves injecting commercial and advertising knowledge into the base LLM. We collected knowledge from advertising systems and produced synthesized data, which were then injected into Hunyuan-1B and Hunyuan-13B models for online and offline scenarios, respectively. For the detailed information of knowledge data, please see the Table~\ref{tab:appendix-1} of Appendix A. The knowledge injection process can be formalized as follows:s

\begin{equation}\label{formula.ki1}
\begin{split}
{\theta}^{'} = h(\theta,K),
\end{split}
\end{equation}
where the comprehensive function $h$ takes the LLM model parameters $\theta$, and the advertising knowledge data $K$ as inputs, and outputs the updated model parameters $\theta^{'}$. Subsequently, the new parameters $\theta^{'}$ are utilized to generate predictions, i.e.,
\begin{equation}\label{formula.ki2}
\begin{split}
y=P(y|x; {\theta}^{'} ).
\end{split}
\end{equation}

\paragraph{Stage 2: Format Fine-Tuning.} 
Building on the LLM enhanced with commercial knowledge, this stage focuses on refining the format of the generated CIs and increasing their diversity. The training data for format fine-tuning is obtained from real-world online data after making necessary format adjustments. For the detailed information of fine-tuning data, please see the Table~\ref{tab:appendix-1} of Appendix A. The generation loss of format fine-tuning is shown as follows: 
\begin{equation}\label{formula.fft}
\begin{split}
L(\theta) = {{1} \over N} \sum\limits_{i=1}^N \sum\limits_{t=1}^{ T_{i} }logp( y_{i,t} | y_{i<t}, x_{i};\theta ),
\end{split}
\end{equation}
where fine-tuning data set is $D= {( x_{i}, y_{i} )}_{i=1}^N$, $x_{i}$ is the input sequence and $y_{i}$ is the target output sequence. The probability $p( \cdot )$ is the probability predicted by the model with parameters $\theta$ base on $x_{i}$ and the previously generated words $y_{i<t}$. 

The customization of LLM significantly enhances its ability to understand and extract the intentions behind ads and user queries. The customized LLM compresses and summarizes ads into a commercial intention space, clustering similar ads.This process enhances diversity by reducing homogeneous retrieval, leading to improved retrieval performance both online and offline.

\subsection{Indexing} 
\label{sec:indexing} 
We use the customized LLM to generate the Commercial Intentions (CIs) of ads and then construct the inverted index of CIs-Ads. 
\paragraph{Commercial Intentions (CIs).} 
CIs are short texts generated by the customized LLM that describe the commercial intentions of users or ads. 
Given a prompt containing ad information (such as ad title, landing page) or a user query, the generation of CIs is formalized as:

\begin{equation}\label{formula.1}
\begin{split}
CIs = &arg \max\limits_{ y^{<1>} ...y^{<b>}} \sum\limits_{t=1}^{T} logP(  y_{t}^{<1>}, y_{t}^{<2>}...\\
&y_{t}^{<b>}| x, y_{t-1}^{<1>}, y_{t-1}^{<2>}...y_{t-1}^{<b>}) ,
\end{split}
\end{equation}
where $y_t^{<i>}$ is the output of the top-$i$ commercial intention at time $t$, $b$ is the beam size, and $T$ is the maximum length of the CIs.

\begin{figure}
  \includegraphics[width=\columnwidth]{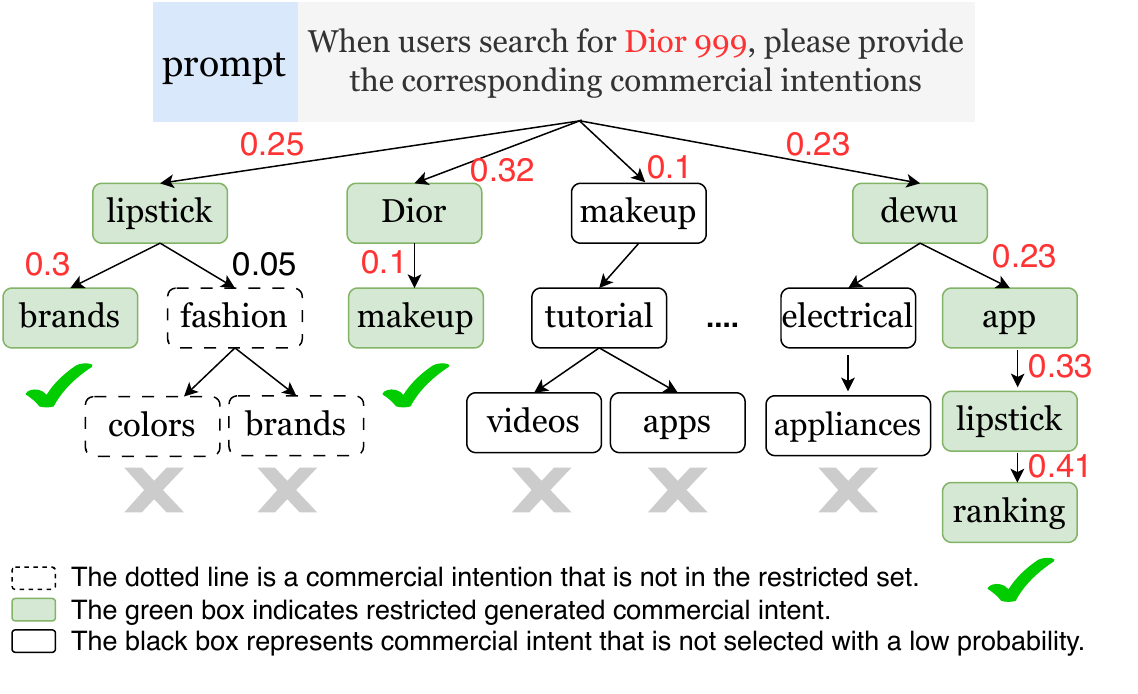}
  \caption{Constrained Beam Search Decoding Process.}
  \label{fig:beam-search}
\end{figure}

\textit{Example.} When the ad pertains to "flowers", RARE not only generates multiple business intents related to flowers—such as buying flowers online, finding a local flower shop, comparing flower prices, ordering flower delivery, and arranging flowers—but also includes intents for occasions like "Mother's Day" and "Valentine's Day". The CIs proposed in RARE can more accurately align with the traffic advertisers want to reach.

\paragraph{Ad Indexing Building.} 
Initially, We use the customized LLM to generate CIs for all ads in the library, based on information such as the ads' titles, landing pages, and delivery materials. Subsequently, we perform operations such as deleting irrelevant CIs and clustering to refine the generated results, resulting in a refined set of approximately 2 million CIs.  For new ads, we employ a constrained beam search technique to generate an average of 30 CIs per ad. This approach ensures that each new ad can be effectively indexed. In addition, we update the CIs set monthly to introduce new products and refine commercial intents.

Utilizing generated CIs to index ads offers numerous advatages, including accurate extraction of both ad content and user intent, as well as high efficiency and robust generalization capabilities. For new ads, our approach performs only simple inference rather than retraining the model.

\subsection{Efficient inference}
\label{sec:reasoning}
Efficient inference is essential for real-time retrieval from millions of candidate sets, as sponsored search advertising has strict requirements on retrieval time. In this section, we mainly introduce efficient decoding methods including constrained decoding and caching technology.

\paragraph{Constrained Beam Search.}  In this work, we employ a constrained beam search algorithm for generating commercial intentions(CIs), ensuring that the model's outputs are confined to a predefined CIs. We have developed a CUDA-based implementation of the constrained beam search and integrated it with the LLM inference process to enable parallel generation of beam-size CIs, thereby enhancing decoding efficiency. Furthermore, we introduced a truncation function within the constrained beam search framework, which allows for the discarding of individual tokens with lower scores to improve the accuracy of the model's output. The specific restriction process is illustrated in Figure~\ref{fig:beam-search}.

\paragraph{Caching Technology.} The search system exhibits a pronounced long-tail effect, where 5\% of the queries account for 60\% of the total query requests. To enhance inference efficiency, we perform offline inference and storage for these high-frequency queries. When a user submits a query, the system first checks the offline cache. If a match is found, the result is returned immediately. If no match is found, the inference service processes the request.

Offline processing is less time-sensitive, allowing us to utilize a large model\textemdash a 13B LLM\textemdash to handle these queries. Online inference has stringent time constraints, typically requiring completion within milliseconds, so a small size 1B model is used. By caching millions of head queries offline, we can reduce online machine consumption by 70\%, which not only decreases the time required for inference but also enhances the quality of CIs for head queries.


\begin{table*}[]
\centering
\begin{tabular}{c|c|lllll}
\toprule
\multicolumn{2}{c|}{\textbf{Method}}                                                           & \multicolumn{1}{c}{\textbf{HR@50}} & \multicolumn{1}{c}{\textbf{HR@100}} & \multicolumn{1}{c}{\textbf{HR@500}} & \multicolumn{1}{c}{\textbf{MAP}} & \multicolumn{1}{c}{\textbf{ACR}} \\ \midrule

{Word-based}                                      & BM25   & 0.0870 & 0.1336 & 0.3807 & 0.1232 & 76.01\% \\ \midrule
\multirow{3}{*}{Semantic-based}   & Bert-small   & 0.0995 & 0.1518 & 0.4311 & 0.1719 & 78.65\% \\
& Bert-base  &  \textbf{0.1038} & 0.1511 & 0.4714 & 0.1739 & 
80.50\% \\
& SimBert-v2-R  &   0.0978 & 0.1428 & 0.3419 & 0.1797 & 81.07\% \\ \midrule
\multirow{2}{*}{\begin{tabular}[c]{@{}c@{}} Generative \\ Retrieval\end{tabular}}  & SimBert-v2-G      &   0.0572 & 0.0792 & 0.1026 & 0.1405 & 43.27\%    \\
   & T5      &  0.0265    &        0.0447  & 0.1130    &  0.1036      &   83.31\%      \\ \midrule
\multirow{4}{*}{\begin{tabular}[c]{@{}c@{}} LLM-based \\ Generative \\ Retrieval\end{tabular}} & Qwen-1.8B   & 0.0527 & 0.0986 & 0.4099 & 0.1168 & 96.13\%\\
& Hunyuan-2B   & 0.0491 & 0.0937 & 0.3904 & 0.1038 &  96.09\% \\
   & DSI  &  0.0258 & 0.0480 & 0.1764 & 0.0745 & \textbf{96.15\%}\\ 
& Substr   &  0.0225 &  0.0341 & 0.0744 & 0.1042 & 96.15\% \\
\midrule

Ours   & \textbf{RARE}               &     \multicolumn{1}{l}{0.0985}                       &         \multicolumn{1}{l}{\textbf{0.1541}}                   &           \multicolumn{1}{l}{\textbf{0.5134}}                 &        \multicolumn{1}{l}{\textbf{0.1845}}                     &      \multicolumn{1}{l}{95.05\%}                       \\ \bottomrule

\end{tabular}
\caption{Comparison of RARE and Baseline Models in Offline Scenarios.}
  \label{tab:eval-res}
\end{table*}

\section{Experiments}

In this section, we primarily introduce our experimental settings, discuss the effect of RARE on offline metrics and online systems, and present the results of ablation studies.

\begin{table*}
  \centering
  \begin{tabular}{c|ccccc}
    \toprule
   {\textbf{\begin{tabular}[c]{@{}c@{}}Real-Time\\ Online Scenarios\end{tabular}}} & \textbf{Consumption} & \textbf{GMV} & \textbf{CTR} & \textbf{\begin{tabular}[c]{@{}c@{}}Shallow \\ Conversions\end{tabular}} & \textbf{\begin{tabular}[c]{@{}c@{}}Deep\\  Conversions\end{tabular}} \\ 
    \midrule
    \verb|WeChat Search|  & {+5.04\%}  & {+6.37\%}  & {+1.28\%} &{+5.29\%} &{+24.77\%}       \\
    \verb|Demand-Side Platform|  & {+7.18\%} &{+5.03\%} &{-} &{+6.85\%} &{+5.93\%}   \\
    \verb|QQ Browser Search|  & {+4.50\%} &{+5.02\%} &{-0.74\%} &{+17.07\%} &{+7.86\%}         \\
     \bottomrule
  \end{tabular}
  \caption{Application of RARE to Real-World Search Systems. Results of Online A/B Testing.}
  \label{tab:ab-testing}
\end{table*}

\subsection{Experimental Settings} 

\paragraph{Training Dataset.} 


To facilitate knowledge injection into the vanilla LLM, we utilized commercial knowledge and synthetic data. The raw data was derived from real online logs and was processed to generate the final synthetic data by having open-source LLMs perform tasks such as query intent mining and ad intent mining.
Format fine-tuning primarily involves the CIs of queries and advertisements. These data are sourced from real online interactions and are combined according to fixed rules. For details, please refer to Table~\ref{tab:appendix-1}. 

\paragraph{Evaluation Dataset.} 
To evaluate the model's effectiveness, we collected pairs of head queries and corresponding clicked ads online over the course of one day in the real-world scenario. After cleaning the data, we obtained 5,000 queries and 150,000 ads to serve as the ground truth, with each query having a maximum of 1,000 ad candidates. 

\paragraph{Baselines.} 
We compare RARE with 10 competitive baselines across 4 major categories, including word-based BM25,  semantic-based BERT, generative-based T5 and LLM-based Qwen, etc.

BM25 segment the query to be calculated into $w_{1}$, $w_{2}$, ..., $w_{n}$, and then calculate the relevance score of each $w_{i}$ and the keyword. Finally, these scores are accumulated to finally get the text similarity calculation result.

\begin{figure}
  \includegraphics[width=\columnwidth]{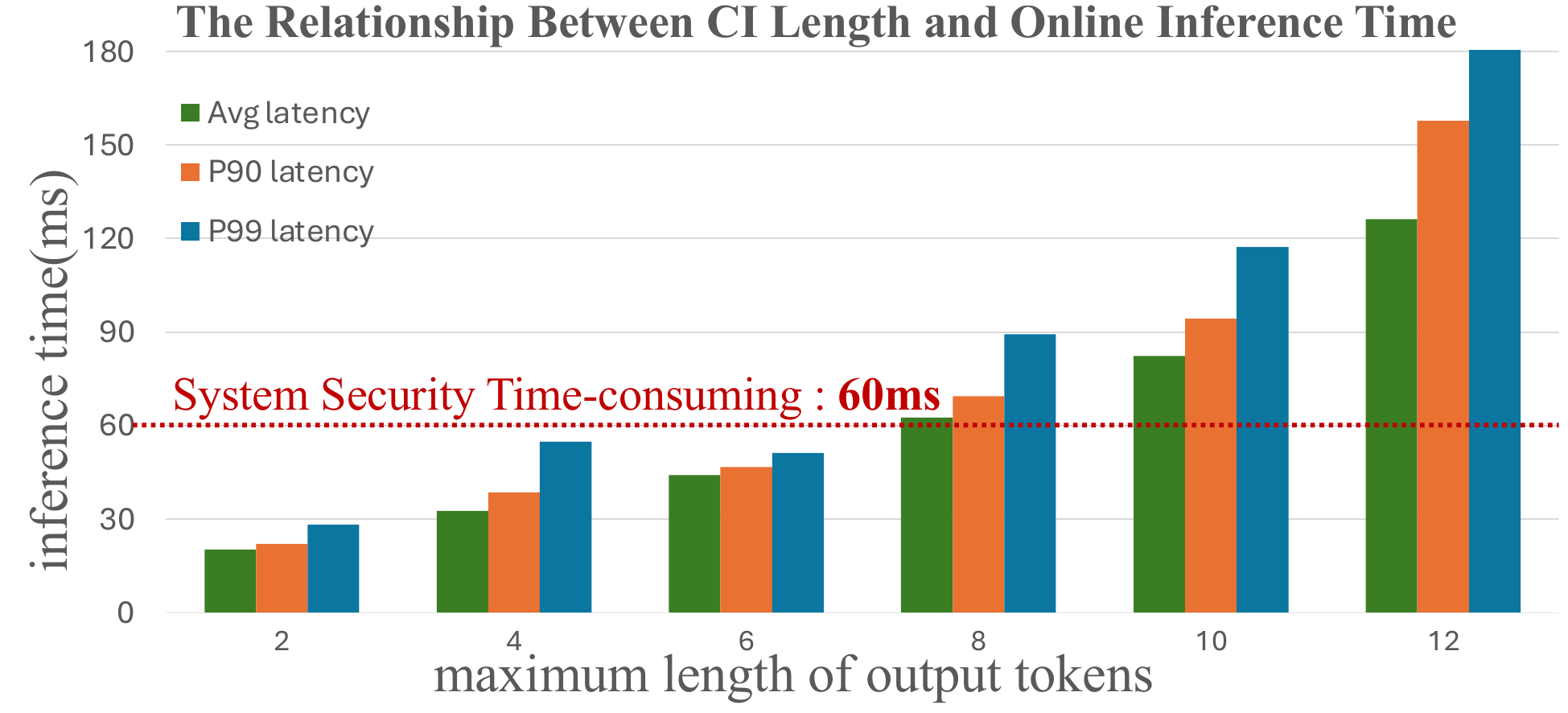}
  \caption{Time Consumption for Different Lengths.}
  \label{fig:time-bs}
\end{figure}

BERT-small employs a 4-layer transformer network with a hidden layer size of 768 and the number of parameters is approximately 52.14M. BERT-base utilizes a 12-layer transformer network with a hidden layer size of 768 and 12 heads.  The number of parameters is about 110M. We used online clicked data as positive examples, and randomly sampled within the batch as negative examples. We trained the BERT using contrastive learning techniques. The trained BERT was used to obtain the embeddings of the query and keywords and then use HNSW~\cite{HNSW:16} to retrieve candidate keywords for the query. SimbBert-v2-R~\cite{jianlinsu} is a model that integrates both generation and retrieval capabilities. It serves as a robust baseline for sentence vectors and can also be utilized for automatic text generation. In our work, we reproduced the Simbert-v2-base\footnote{https://github.com/ZhuiyiTechnology/roformer-sim} model and trained it on millions of online click query-keyword pairs. This resulted in two specialized versions: Simbert-v2-G, designed for keyword generation, and Simbert-v2-R, intended for calculating keyword sentence vectors. We also reproduced T5-base\footnote{https://github.com/bojone/t5\_in\_bert4keras}, a strong baseline for generative recall, and fine-tuned it on a large number of online click queries and keywords.

DSI is a typical method that employs semantic ID-based retrieval. Initially, we fine-tune the HunYuan-1B model to learn the correspondence between ads and their respective IDs. Subsequently, we input the query along with the IDs corresponding to the clicked ads into the HunYuan model for further training. Qwen-1.8B and Hunyuan-2B are models with the same scale of parameters as RARE. We incorporated format fine-tuning into the Qwen 1.8B and Hunyuan-2B models to ensure that they generate outputs exclusively focused on CIs, without including any additional information. A LLM without constrained decoding may generate CI without corresponding advertisements. To address this issue, we employ HNSW retrieval to find the most similar CI within the library of CIs, using it as the final result for the CI generated by the LLM.

\paragraph{Evaluation Metrics.} We use ACR~\cite{fan2019mobius}, Hit Ratio (HR@K)~\cite{alsini2020hit} and Mean Average Precision (MAP)~\cite{cormack2006statistical} to evaluate the effectiveness of RARE. 

\textit{Ad Coverage Rate (ACR)} in ad retrieval means coverage, which is the proportion of requests with ad recall. As shown in Formula~\ref{formula.0}, Ad Pave View (AdPV) is the number of requests with ad recall, and Pave View (PV) is the number of requests.

\begin{equation}\label{formula.0}
{ACR = AdPV/PV}
\end{equation}

\textit{Hit Ratio (HR@K)} is shown in Formula ~\ref{formula.2}, where Ground Truth (GT) represents set of candidate ads, and Hits@K represents the number of relevant ads within the top-K retrieved candidates that belong to the ground truth set.

\begin{equation}\label{formula.2}
{HR@K =  \frac{Hits@K}{|GT|} }
\end{equation}

\textit{Mean Average Precision(MAP)} is the average of Average Precision (AP) of all queries (Q), as shown in formula~\ref{formula.4}.

\begin{equation}\label{formula.4}
MAP =  \frac{ \sum_{q \in Q} A P_{q}  }{Q}
\end{equation}

\textit{Average Precision (AP)} is shown in formula~\ref{formula.3}, where $\Omega_{q} $ represents the ground-truth results, $p_{qj} $ represents the position of $ad_{j}$ in the generated list, and $p_{qj} <  p_{qi}$ means that $ad_{j}$ ranks before $ad_{i}$ in the generated list.

\begin{equation}\label{formula.3}
AP_{q}  =  \frac{1}{ \Omega_{q} }  \sum\limits_{i \in  \Omega_{q} }  \frac{ \sum_{j \in \Omega_{q}} h( p_{qj} <  p_{qi}) +1     }{ p_{qi} }
\end{equation}

\begin{table*}
  \centering
  \begin{tabular}{c|ccccc}
        \toprule
        \textbf{Method} &  \textbf{HR@500} & \textbf{MAP} & \textbf{Recall} & \textbf{Avg CIs} & \textbf{Accuracy} \\
        \midrule
        \textbf{w/o. KI}  & {0.1706}  & {0.1540} & {59.51\%} & {22.78} & {90.4\%}  \\
        \textbf{ w/o. CBS}  & 0.1868  & 0.1687 & 67.12\% & 4.84 & 95.2\%  \\
         \textbf{w/o. CBS \& KI}   & 0.1562  & 0.1592 & 48.28\% & 9.09 & 94.5\%  \\
         \textbf{RARE} & {\textbf{0.5134}}  & {\textbf{0.1845}} & {\textbf{95.05\%}} & {\textbf{74.49}} & {\textbf{96.5\%}}  \\
         \bottomrule
      \end{tabular}
  \caption{Ablation Studies on RARE.}
  \label{tab:ab-stu}
\end{table*}

\paragraph{Implementation Details.} We utilize Hunyuan as the backbone, with parameters including 1B-Dense-SFT and 13B-Dense-SFT. For the offline cache, we employ a 13B model with a beam size of 256, a temperature of 0.8, and a maximum output length of 6. For online inference, we use a 1B model with a beam size of 50, a temperature of 0.7, and a maximum output length of 4 to ensure that inference latency remains within 60 milliseconds. RARE will assign appropriate CIs to newly added advertisements and products within the existing CI set, and will update the CIs-Ads index on an hourly basis. The entire CIs set is updated monthly, allowing new products to receive more fine-grained and accurate CIs. Additionally, we periodically inject new commercial information into the LLM, such as new brand names and product details, to ensure its knowledge remains up to date.

\subsection{Experimental Results}
\textbf{Offline Evaluation.} We compared RARE with 10 retrieval methods in 4 categories on the industrial evaluation dataset. Results are shown in Table~\ref{tab:eval-res}. The RARE model excels in HR@500 and MAP while maintaining a high ACR, demonstrating its ability to understand user search intent and optimize ad delivery. Notably, it achieves a ACR exceeding 90\%, and its high HR@500 metric confirms its strong capacity to retrieve commercially valuable ads. This synergy indicates the model's success in balancing user intent comprehension with commercial value-driven ad retrieval.

\paragraph{Ablation Study.} 
We conducted two types of ablation studies to investigate the contribution of each component. First, table~\ref{tab:ab-stu} displays the results of RARE on ad retrieval under various settings. w/o. KI refers to RARE without knowledge injection. Its recall rate is only 59.51\%, significantly lower than RARE's 95.05\%. This demonstrates that without knowledge injection, the LLM struggles to understand intents of user queries and ads. w/o. CBS refers to RARE without constrained beam search. Its average number of CIs is only 4.84, significantly lower than RARE's 74.49. This indicates that constrained beam search can substantially increase the diversity of commercial intents generated by the LLM. w/o.CBS \& KI refers to RARE without both constrained beam search and knowledge injection. It is evident that its HR@500, MAP, and ACR metrics are the lowest among the compared methods. Second, table~\ref{tab:ab_detail} in Appendix~\ref{qualitative_abl} presents a qualitative analysis of each component's contribution to RARE through a case study.

\begin{figure}
  \includegraphics[width=\columnwidth]{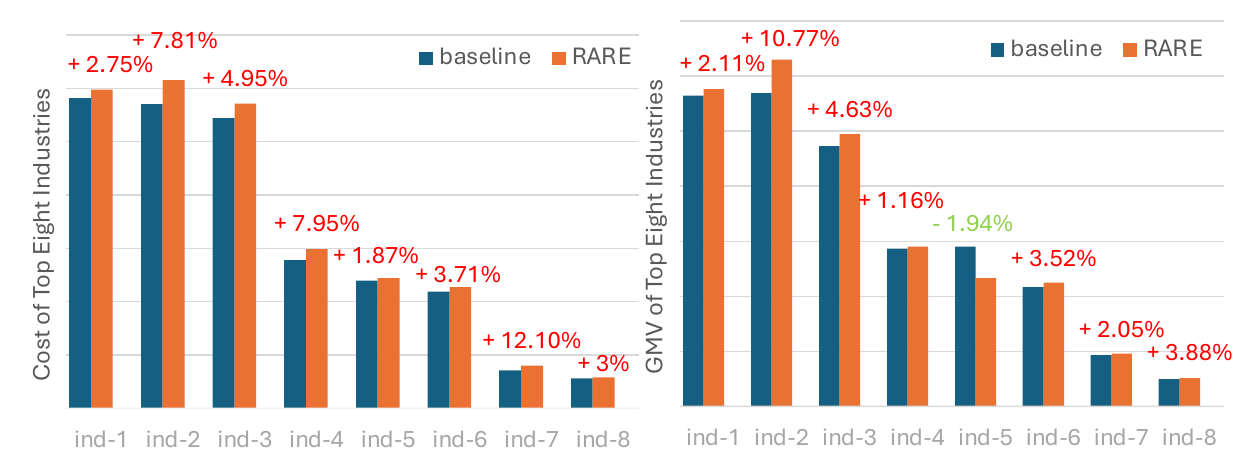}
  \caption{RARE Outperforms Online Benchmark Models Across Major Real-World Industries.}
  \label{fig:top-info}
\end{figure}

\paragraph{Online A/B Testing.} We apply RARE to three different Tencent online retrieval scenarios (with billions of daily requests): WeChat Search (WTS), Demand-Side Platform (DSP) and QQ Browser Search (QBS). During a one-month A/B testing experiment with a 20\% user sample, we observed significant benefits across multiple scenarios, including increased system revenue, enhanced user experience, and boosted advertiser conversions. Take WTS as an example, we achieved a 5.04\% increase in consumption (cost), a 6.37\% increase in GMV, a 1.28\% increase in CTR and a 5.29\%  increase in shallow conversions. Significant improvements of CTR and conversions demonstrate that RARE can effectively understand user intent and deliver high-quality ads. 
The evaluation of RARE across eight popular real-world industries, as shown in Figure~\ref{fig:top-info}, further demonstrates its effectiveness in various scenarios.

\paragraph{Online Inference Support.} We examined the time consumption of various output lengths during real-time online inference, with results shown in figure~\ref{fig:time-bs}. Our CIs have an average token count of 3, ensuring that online real-time inference meets safety thresholds. To facilitate online inference, we developed a specialized GPU cluster with hundreds of L40, achieving effective load balancing and peak GPU utilization rates up to 90\%. We quantized the well-trained model to FP8 precision, enabling each L40 to handle about 30 Queries Per Second. Efficient caching techniques increased the cache hit rate to approximately 65\% for head queries. These supports enhanced generation quality while reducing computational costs notably.

\section{Conclusion}

In this paper, we propose a LLM-generative \textbf{R}eal-time \textbf{A}d \textbf{RE}trieval called RARE. This framework utilizes commercial intentions (CIs) as semantic representation that retrieve ads directly for querys. To mine deeper intentions of ads and users, we inject commercial knowledge and conduct format fine-tuning on vanilla LLM to obtain the customized LLM.
Besides, we employs constrained decoding, which allows the model to generate CIs from a fixed set in parallel.
The proposed architecture enables real-time generation and retrieval from a library containing tens of millions of ads. Evaluations on offline data and online A/B testing indicate that our architecture achieves state-of-the-art (SOTA) advertising retrieval performance, while substantially improving search system revenue, user experience and advertiser conversion.

\section*{Limitations} We briefly outline limitations of our work. The end-to-end generation architecture proposed in this paper primarily facilitates the generation process from query/ad to commercial intention, while the correlation between query and ad is managed by downstream processes. In future work, we aim to integrate correlation assessment into LLMs, thereby empowering the model to evaluate the pertinence between prompts and commercial intents concurrently with the generation phase. We anticipate that this integration of generative and discriminative capabilities will significantly augment the efficacy of the generation process.



\bibliography{custom}
\newpage
\appendix

\section{Fine-tuning Data}
\label{appendixA}
In this section, we mainly introduce the details of fine-tuning data. Fine-tuning of customized LLM mainly includes two stages, namely knowledge injection and format fine-tuning. The fine-tuning data of knowledge injection mainly includes query intent mining, advertising intent mining and advertising words buying. We input prompts containing advertisement and user information into open-source LLMs (e.g., ChatGPT) to obtain outputs that include rich reasoning processes and guidance information, which are then injected into the Hunyuan model as knowledge data. Injecting a large amount of data (on the order of hundreds of thousands or millions) during the fine-tuning phase can cause LLMs to lose their general knowledge and reasoning capabilities. Therefore, in this phase of fine-tuning, we selected only 2,000 instances for each task. Table~\ref{tab:appendix-1} shows the fine-tuning data of these two stages in detail.

\section{Related Work}

\textbf{Beam Search.} As a decoding strategy for heuristic search, beam search has been widely used in many works. For example, DSI uses beam search to generate a sorted list of candidate documents, and Tiger uses beam search to generate multiple candidate product IDs at once. As early as a few years ago, the combination of seq2seq and constrained Beam Search has achieved a win-win effect and efficiency in entity linking and document retrieval. For example, GENRE (\citealp{genre:98}) applied constrained Beam Search to document retrieval tasks and achieved SOTA.

\textbf{Query-kwds-ads Architecture.}   Traditional query-kwds-ads approaches suffer from two critical drawbacks: (1) Keywords are manually selected by advertisers, resulting in varying quality and potential issues of either being too broad or too narrow, leading to inefficient traffic matching. (2) Advertisers often purchase a large number of keywords, which hampers the efficiency of ad retrieval after keyword inversion, imposing a significant burden on the system. In contrast, CIs are generated by a domain knowledge-injected LLM, enabling them to better represent the intentions of advertisers and achieve more precise matching with relevant traffic. This not only brings economic benefits but also ensures the long-term healthy operation of the system.

\textbf{Encoder-based LLMs Retriever} Encoder-based retrievers leverage the semantic capabilities of LLMs to obtain text embedding~\cite{hou2024bridging}. For instance, cpt-text~\cite{cpt-txt:27} uses contrastive learning to train GPT-3~\cite{gpt-3:25} from scratch, generating high-quality embedding of text. GTR~\cite{Gtr:22} utilizes the T5~\cite{T5:26} model, fine-tuning it to derive text vector representations. NoteLLM~\cite{Notellm:21} achieves text-image embeddings by incorporating additional training tasks and modifying the LLM model structure. 

\begin{table*}
  \centering
  \begin{tabular}{m{4cm}<{\centering}m{5.5cm}m{4.5cm}m{1cm}<{\centering}}
    \hline
    \textbf{Task Name} &  \textbf{Prompt} &  \textbf{Output} & \textbf{Data Size} \\
    \hline
    \multirow{5}{*}{\begin{tabular}[c]{@{}c@{}} \textbf{Knowledge Injection} \\ Task1. query intent mining \end{tabular}}  & \begin{CJK}{UTF8}{gbsn} \small{As a search advertising system, given the user query = xxx, identify the commercial intentions based on the user's search characteristics. Provide ten commercial intentions for each category from different intents, and finally, compile all the results into a single line separated by commas, step by step. } \end{CJK} & \begin{CJK}{UTF8}{gbsn}\small{\textbf{Output from open source LLM models (e.g. ChatGPT):}Sure, let's break down the process step by step to identify commercial intentions based on the user query "xxx".        \textit{Step 1: Identify Categories}...       \textit{Step 2: Identify Commercial Intentions for Each Category}...       \textit{Step 3:Compile All Results into a Single Line}... }\end{CJK} & 2000 \\ \hline
    \multirow{5}{*}{\begin{tabular}[c]{@{}c@{}} \textbf{Knowledge Injection} \\ Task2. ad intent mining \end{tabular}}  & \begin{CJK}{UTF8}{gbsn} \small{ As an advertiser, given the ad title = xxx, identify the commercial intentions for ad based on search characteristics. Provide ten commercial intentions for each category from different intents, and finally, compile all the results into a single line separated by commas, step by step.  } \end{CJK}  & \begin{CJK}{UTF8}{gbsn}\small{\textbf{Output from open source LLM models (e.g. ChatGPT): }Sure, let's break down the process step by step to identify commercial intentions based on the ad title "xxx" \textit{Step 1: Identify Categories...}  \textit{Step 2: Identify Commercial Intentions for Each Category...} \textit{Step 3: Compile All Results into a Single Line....}  }\end{CJK} & 2000 \\  \hline
    \multirow{5}{*}{\begin{tabular}[c]{@{}c@{}} \textbf{Knowledge Injection} \\ Task3. ad generation  \end{tabular}}   & \begin{CJK}{UTF8}{gbsn} \small{As a search advertising system, your task is to recommend suitable ads for user queries. To meet the user's potential intent and ensure the relevance between the query and the ad title, you should provide a diverse range of ads. Given the query = xxx, please generate ten appropriate ad titles for each intent based on search characteristics, step by step.}\end{CJK}  & \begin{CJK}{UTF8}{gbsn}\small{\textbf{Output from open source LLM models (e.g. ChatGPT):} Sure, let's break down the process step by step to generate ten appropriate ad titles for each intent based on the user query "xxx".  \textit{Step 1: Identify Intents...}  \textit{Step 2: Generate Ad Titles for Each Intent...}}\end{CJK} &  2000 \\ \hline
    \multirow{5}{*}{\begin{tabular}[c]{@{}c@{}} \textbf{Knowledge Injection} \\ Task4. query understanding \end{tabular}}   & \begin{CJK}{UTF8}{gbsn} \small{As a search advertising system, your task is to rewrite and expand the user's query to ensure it meets the diverse needs of users. The rewritten terms should be as varied as possible. For instance, if the query is "SF Express," the expanded terms could include "cross-border logistics", "express shipping", "express delivery", "express service" and "logistics warehousing". Following this example, when the query is = xxx, provide the expanded terms that meet these requirements, step by step.} \end{CJK}  & \begin{CJK}{UTF8}{gbsn}\small{\textbf{Output from open source LLM models (e.g. ChatGPT):} ure, let's break down the process step by step to rewrite and expand the user's query "xxx" to ensure it meets the diverse needs of users. The expanded terms should cover a wide range of related topics and intents. \textit{Step 1: Identify Core Concepts...} \textit{Step 2: Identify Related Categories and Intents...} \textit{Step 3: Generate Expanded Terms for Each Category...} \textit{Step 4: Compile All Expanded Terms...}} \end{CJK}&  2000 \\ \hline
    \multirow{3}{*}{\begin{tabular}[c]{@{}c@{}} \textbf{Format Fine-tuning} \\ Task1. query generates CIs  \end{tabular}}  & \begin{CJK}{UTF8}{gbsn} \small{As a search advertising system, given a user query of xxx, please provide a broader range of commercial intentions that still align with the user's potential intentions.} \end{CJK}   & \begin{CJK}{UTF8}{gbsn} \small{\textbf{From real world online data:} commercial intention 1;commercial intention 2;commercial intention 3.....} \end{CJK}&  2000\\ \hline
    \multirow{3}{*}{\begin{tabular}[c]{@{}c@{}} \textbf{Format Fine-tuning} \\ Task2. ad generates CIs  \end{tabular}}  & \begin{CJK}{UTF8}{gbsn} \small{As a search advertiser, given the ad title = xxx, please identify the commercial intentions of the ad based on the characteristics of the search and various user intentions.} \end{CJK}  & \begin{CJK}{UTF8}{gbsn} \small{\textbf{From real world online data:} commercial intention 1;commercial intention 2;commercial intention 3.....} \end{CJK} &  2000\\
     \hline
  \end{tabular}
  \caption{Details of fine-tuning data for customized LLM.}
  \label{tab:appendix-1}
\end{table*}

\section{Qualitative Analysis}
\label{qualitative_abl}
Table~\ref{tab:ab_detail} provides an intuitive example to analyze the role of each component in RARE. we can observe the following: (1) Zero-shot LLM lacks a reasoning process for prompts, relying mainly on the surface-level understanding of queries, which results in numerous poor cases. (2) Knowledge injection stage teaches LLM how to reason, enabling it to analyze the query in the first step, provide commercial intentions in the second step, and judge relevance in the third step. However, this approach lacks diversity. (3) Format fine-tuning stage instructs the model to follow rules while generating more diverse results. Since most of the training data in this stage comes from online data, the commercial intentions are relatively rich. However, due to the free generation process, the number of commercial intentions is limited. (4) Incorporating constrained decoding can increase the number of commercial intentions while ensuring relevance and diversity.

\begin{table*}{}
  \centering
  \begin{tabular}{p{2.0\columnwidth}}
    \hline
    \textbf{prompt}: When users search for \textbf{shantytown housing}, please provide the corresponding commercial intentions. \\
    \hline 
    \textbf{vanilla llm:} (A total of \textcolor{red}{26 commercial intentions}, but some of them are \textcolor{blue}{poorly relevant}) \\
    commercial intentions: shantytown, \textcolor{red}{demolition}, \textcolor{red}{housing supply}, \textcolor{blue}{price},  \textcolor{blue}{user feedback}, \textcolor{blue}{case sharing}..., etc. \\
    \hline
    \textbf{w/o. CBS\&FFT } (A total of \textcolor{red}{9 commercial intentions}, with \textcolor{red}{good correlation} and \textcolor{blue}{insufficient diversity}.) \\
    \textcolor{red}{Step 1: Mining user query intent} \\
    Shantytown reconstruction is a government-led urban renewal project involving demolition.... \\
    \textcolor{red}{Step 2: Generate commercial intentions} \\
    Shantytown reconstruction policy, demolition compensation standards.....  \\
    \textcolor{red}{Step3: Ensure commercial intentions are relevant to the query} \\
    commercial intentions such as shantytown reconstruction policy, Demolition compensation standards... \\
    \textcolor{red}{Final result:} shantytown reconstruction policy, demolition compensation standards.....  \\
    \hline
    \textbf{w/o. CBS} (A total of \textcolor{red}{12 commercial intentions}, with \textcolor{red}{good correlation and diversity}, but \textcolor{blue}{a small number}) \\
    Anjuke rents a house, second-hand housing, house hunting...  \\
    \hline
    \textbf{RARE} (A total of with \textcolor{red}{142 commercial intentions}, with with \textcolor{red}{good correlation, good diversity and large number}) \\
    Anjuke house hunting, demolition compensation, new house decoration, renovation of old houses, public housing application...... \\
    \hline
  \end{tabular}
  \caption{Commercial intention generation effects based on different fine-tuning methods.}
  \label{tab:ab_detail}
\end{table*}

\end{document}